\documentclass[pra,showpacs,twocolumn]{revtex4}
\usepackage{amsfonts}
\usepackage{amsmath}
\usepackage{amssymb}
\usepackage{graphicx}
\renewcommand{\Re}{\mathop{\mathrm{Re}}}
\renewcommand{\Im}{\mathop{\mathrm{Im}}}

\begin{document}

\title{Dynamical generation of phase-squeezed states in a two-component
Bose-Einstein condensates}

\author{G. R. Jin, Y. An, T. Yan, and Z. S. Lu}
\affiliation{Department of Physics, Beijing Jiaotong University,
Beijing 100044, China}

\begin{abstract}
As an ``input" state of a linear (Mach-Zehnder or Ramsey)
interferometer, phase-squeezed state proposed by Berry and Wiseman
exhibits the best sensitivity approaching to the Heisenberg limit
[Phys. Rev. Lett. \textbf{85}, 5098 (2000)]. Similar with the Berry
and Wiseman's state, we find that two kinds of phase-squeezed states
can be generated dynamically with atomic Bose-Einstein condensates
confined in a symmetric double-well potential, which shows the
squeezing along spin operator $\hat{S}_y$ and the anti-squeezing
along $\hat{S}_z$, leading to sub-shot-noise phase estimation.
\end{abstract}

 \maketitle


\section{Introduction}

Recently, atom interferometer with a trapped Bose-Einstein
condensates (BEC) has attracted much attention due to its potential
applications in quantum information and quantum
metrology~\cite{Schumm,Shin,Smerzi06,Grond,Cronin,Gross,Riedel}. In
particular, a two-component BEC with the `one-axis twisting' (OAT)
interaction $\chi\hat{S}_{z}^{2}$ is widely believed to be a useful
resource for preparing the OAT-type spin squeezed
state~\cite{Kitagawa,Sorensen,Wang,Jin09}, as well as many-partite
entangled state~\cite{Molmer,Rey08}. Using maximally entangled
(i.e., the N00N) state rather than product coherent spin states
(CSS), phase sensitivity of a linear (Mach-Zehnder, or equivalently
Ramsey) interferometer can be improved from the so-called shot-noise
limit (SNL) $\Delta\Phi\sim 1/\sqrt{N}$ to the Heisenberg limit (HL)
$\sim 1/N$~\cite{Wineland1,Wineland2,Mitchell,Giovannetti}, where
$N$ is total particle number. However, it is very difficult to
create the N00N state.

As a special case of spin squeezed state, Gaussian number-squeezed
state (NSS) has been demonstrated in several
experiments~\cite{Orzel,Strabley,Esteve,Chuu,Jo,Vuletic}, which
exhibits sub-Poissonian atom number distribution~\cite{Orzel,Chuu}
and long coherence time~\cite{Jo} due to the spin squeezing of
relative-number operator $\hat{S}_z$. These features manifest it as
a promising candidate for the Heisenberg-limited
metrology~\cite{Uys,Huang}. Dynamical generation of the NSS has been
proposed in a two-component BEC by exploiting the interplay of
nonlinear interaction and atomic tunneling~\cite{Law,Jin07}. It was
found that the variance of $\hat{S}_z$ has an exact relationship
with the visibility of the Ramsey signal (i.e., the phase
coherence)~\cite{Jin08,Jin10}.

In this paper, we further investigate dynamical generation of
phase-squeezed states (PSS) with atomic BEC confined in a symmetric
double well. We focus on a symmetric input state of a linear
interferometer, with which we derive easily the phase sensitivity
and its relation with spin squeezing parameter. Two experimentally
realizable schemes are proposed to generate the PSS. Similarly with
the Berry and Wiseman's (BW) state~\cite{Berry}, we find that it
exhibits the squeezing of the spin operator $\hat{S}_{y}$ and the
anti-squeezing of $\hat{S}_{z}$. Numerically, we calculate power
rules of the optimal squeezing with particle number $N$. Although
the achievable squeezing is worse than that of the BW state,
dynamically generated PSS are still useful to gain sub-shot-noise
limited phase sensitivity~\cite{Caves}.

This paper is arranged as follows. In Sec.~II we review quantum
metrology protocol based upon the linear interferometer. For a
symmetric input state, we derive the relation between phase
sensitivity and the spin squeezing parameter~\cite{Wineland1}. In
Sec.~III, we investigate general features of the optimal
phase-squeezed state proposed by Berry and Wiseman~\cite{Berry}. Two
routes for creating the PSS are proposed with the two-component
BECs, which utilize a cooperation between the nonlinear interaction
and the Josephson tunneling of ultra-cold atoms confined in a
symmetric double-well potential. Scaling rules of the optimal
coupling and the resultant spin squeezing are obtained to compare
with the BW state. Finally, we conclude in Sec.~IV with main results
of our work.

\section{A symmetric input state of Linear interferometer}

Let us firstly review the achievable sensitivity of a linear
interferometer, which can be described by multiple rotations to the
input state~\cite{Sanders}, namely
\begin{equation}
|\Psi \rangle_{\text{out}}=e^{-i\frac{\pi}{2}\hat{S}_{x}}e^{i\Phi
\hat{S}_{z}}e^{i\frac{\pi}{2}\hat{S}_{x}}|\Psi\rangle
_{\text{in}}=e^{-i\Phi\hat{S}_{y}}|\Psi\rangle_{\text{in}},
\label{MZI}
\end{equation}%
where the three unitary transformations (from left to right)
represent the output 50:50 beam splitter (or equivalently a
$\pi/2$-pulse), the linear phase shifter, and the input beam
splitter, respectively. In the Schwinger representation, the angular
momentum operator $\vec{S}=(\hat{S}_x, \hat{S}_y, \hat{S}_z)$ with
$\hat{S}_{x}=\frac{1}{2}(\hat{a}_{L}^{\dag}\hat{a}_{R}+\hat{a}_{R}^{\dag}\hat{a}_{L})$,
$\hat{S}_{y}=\frac{1}{2i}(\hat{a}_{L}^{\dag}\hat{a}_{R}-\hat{a}_{R}^{\dag}\hat{a}_{L})$,
and
$\hat{S}_{z}=\frac{1}{2}(\hat{a}_{L}^{\dag}\hat{a}_{L}-\hat{a}_{R}^{\dag}\hat{a}_{R})$,
with $\hat{a}_{\mu}$ denoting the annihilation operator for two
Bosonic modes $\mu=L$, $R$. Via a detection of $\hat{S}_{z}$ (i.e.,
the light-intensity difference or population imbalance) at the
output ports, the dimensionless phase shift $\Phi$ could be
estimated with its precision quantified by
\begin{equation}
\Delta\Phi=\frac{(\Delta \hat{S}_{z})_{\text{out}}}{\left\vert
d\langle\hat{S}_{z}\rangle_{\text{out}}/d\Phi\right\vert},
\label{sensitivity}
\end{equation}%
where $\langle\hat{A}\rangle_{\text{out}}$ and $(\Delta
\hat{A})_{\text{out}}=[\langle\hat{A}^{2}\rangle_{\text{out}}-\langle\hat{A}\rangle_{\text{out}}^{2}]^{1/2}$
denote the expectation value and the variance of an operator
$\hat{A}$ with respect to the output state $|\Psi
\rangle_{\text{out}}$.

In terms of common eigenstates of $\hat{S}^2$ and $\hat{S}_{z}$,
$|s,
n\rangle=(a_{L}^{\dagger})^{s+n}(a_{R}^{\dagger})^{s-n}|0\rangle
/\sqrt{(s+n)!(s-n)!}\equiv|s+n, s-n\rangle_{\text{L,R}}$, symmetric
input state of a linear interferometer can be expanded as
\begin{equation}
|\Psi\rangle_{\text{in}}=\sum\limits_{n=-s}^{s}c_{n}\left\vert s,
n\right\rangle,\label{symmetric}
\end{equation}
where $s=N/2$ and the probability amplitudes $c_{-n}=c_{n}$. It is
easy to prove that the mean spin
$\langle\vec{S}\rangle_{\text{in}}=(\langle
\hat{S}_{x}\rangle_{\text{in}}, \langle\hat{S}_{y}\rangle
_{\text{in}}, \langle\hat{S}_{z}\rangle_{\text{in}})$ is oriented in
the $x$ direction. Specially, the $z$th spin component
$\langle\hat{S}_{z}\rangle_{\text{in}}=0$ due to $c_{-n}=c_{n}$. In
addition, expectation value of the ladder operator
$\hat{S}_{+}=(\hat{S}_{-})^{\dagger}=\hat{a}_{L}^{\dag}\hat{a}_{R}$
is given by (even $N$ case)
\begin{equation*}
\langle\hat{S}_{+}\rangle_{\text{in}}=\sum\limits_{n=-s}^{s-1}X_{-n}c_{n}c_{n+1}^{\ast
}=2\sum\limits_{n=0}^{s-1}X_{-n}\Re\left[c_{n}c_{n+1}^{\ast}\right],
\end{equation*}%
where $X_{n}=(s+n)^{1/2}(s-n+1)^{1/2}$, satisfying $X_{\mp n}=X_{\pm
n+1}$. Due to real $\langle\hat{S}_{+}\rangle_{\text{in}}$, we have
$\langle\hat{S}_{y}\rangle_{\text{in}}=\Im\langle\hat{S}_{+}\rangle_{\text{in}}=0$
and
$\langle\hat{S}_{x}\rangle_{\text{in}}=\Re\langle\hat{S}_{+}\rangle_{\text{in}}\neq
0$. Similarly, we can prove that
$\langle\hat{S}_{x}\hat{S}_{y}+\hat{S}_{y}\hat{S}_{x}\rangle
_{\text{in}}=\Im\langle\hat{S}_{+}^{2}\rangle_{\text{in}}=0$ and
$\langle
\hat{S}_{x}\hat{S}_{z}+\hat{S}_{z}\hat{S}_{x}\rangle_{\text{in}}=\Re
\langle\hat{S}_{+}(2\hat{S}_{z}+1)\rangle_{\text{in}}=0$.

The above results, valid also for odd $N$ case, enable us to derive
the phase sensitivity $\Delta\Phi$ in a simple way. For instance, we
get immediately the output signal $\langle\hat{S}_{z}\rangle
_{\text{out}}=-\langle\hat{S}_{x}\rangle_{\text{in}}\sin\Phi$ and
the variance $(\Delta \hat{S}_{z})_{\text{out}}=[(\Delta
\hat{S}_{z})_{\text{in}}^{2}\cos^{2}\Phi+(\Delta
\hat{S}_{x})_{\text{in}}^{2}\sin^{2}\Phi]^{1/2}$~\cite{Huang}, where
we have used the relation: $e^{i\Phi
\hat{S}_{y}}(\hat{S}_{z})^{k}e^{-i\Phi\hat{S}_{y}}=(\hat{S}_{z}\cos\Phi-\hat{S}_{x}\sin\Phi)^{k}$
with $k=1$, $2$. Inserting them into Eq.~(\ref{sensitivity}), we
arrive at
\begin{equation*}
\Delta\Phi=\frac{\sqrt{(\Delta\hat{S}_{z})_{\text{in}}^{2}+(\Delta
\hat{S}_{x})_{\text{in}}^{2}\tan^{2}\Phi}}{\left\vert\langle
\hat{S}_{x}\rangle_{\text{in}}\right\vert},
\end{equation*}%
which is minimized for the phase shift $\Phi=0$, with the best
sensitivity
$(\Delta\Phi)_{0}=(\Delta\hat{S}_{z})_{\text{in}}/|\langle\hat{S}_{x}\rangle_{\text{in}}|$.
Here, the subscript \textquotedblleft $0$" denotes the achievable
sensitivity for $\Phi=0$. Following~Wineland {\it et
al}.~\cite{Wineland1}, we introduce the squeezing parameter
\begin{equation}
\zeta_{\text{S}}=\frac{(\Delta \Phi)_{0}}{(\Delta
\Phi)_{\text{css}}}=\frac{\sqrt{N}(\Delta
\hat{S}_{z})_{\text{in}}}{|\langle\hat{S}_{x}\rangle_{\text{in}}|},
\label{squeezing}
\end{equation}
where $(\Delta\Phi)_{\text{css}}=1/\sqrt{N}$, known as the
shot-noise limit, represents the best sensitivity with product
coherent spin states~\cite{CSS}:
\begin{equation}
|s, \pm s\rangle_{x}=e^{-i\frac{\pi}{2}\hat{S}_{y}}|s, \pm s\rangle.
\label{initial state}
\end{equation}
Note that the mean spin of $|s, \pm s\rangle_{x}$ is also parallel
with the $x$-axis due to $\hat{S}_x|s, \pm s\rangle_{x}=\pm s|s, \pm
s\rangle_{x}$. In addition, the two CSS satisfy minimal-uncertainty
relationship~\cite{CSS}: $(\Delta\hat{S}_{z})^2=(\Delta
\hat{S}_{y})^2=|\langle\hat{S}_{x}\rangle|/2$, with the length of
the mean spin $|\langle\hat{S}_{x}\rangle|=s$. Therefore, as an
input of linear interferometer, the coherent spin states lead to
phase estimation limited by the shot noise $1/\sqrt{N}$ (i.e.,
$\zeta_{\text{S}}=1$).

How to improve the sensitivity beyond the SNL (i.e.,
$\zeta_{\text{S}}<1$) is one of the main subjects of quantum
metrology~\cite{Caves,Sanders}. It has been found that several
nonclassical states can reach the Heisenberg-limit
$(\Delta\Phi)_{0}\propto1/N$, such as the N00N state
$\frac{1}{\sqrt{2}}(|s, s\rangle+|s,
-s\rangle)$~\cite{Wineland1,Wineland2,Mitchell,Giovannetti}, and the
twin-Fock state $|s, 0\rangle$ ($=|N/2, N/2\rangle_{\text{L,R}}$ for
even $N$)~\cite{Holland}. However, experimental realization of these
states is not an easy task. In this paper, we present experimentally
available schemes for preparing the phase-squeezed state and its
application in the quantum metrology.

\section{Dynamical generation of phase-squeezed state}

\begin{figure}[phtb]
\begin{center}
\includegraphics[width=2.8in]{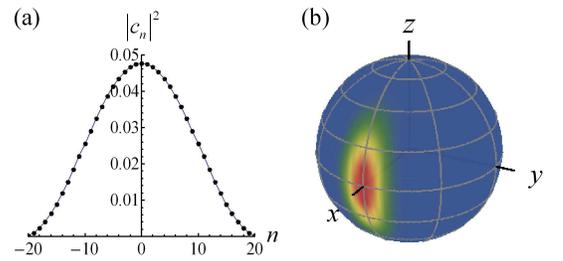}
\caption{(Color online) (a) Probability distribution $|c_n|^2$ for
the Berry and Wiseman's state, with $c_n$ given by
Eq.~(\ref{opPSS}); (b) Quasi-probability distribution $Q(\theta,
\phi)=|\langle\theta, \phi|\Psi\rangle_{\text{BW}}|^2$ on a Bloch
sphere, for $s=N/2=20$ and the coherent spin state $|\theta,
\phi\rangle$ given by Eq.~(\ref{CSS}).} \label{fig1}
\end{center}
\end{figure}

Previously, Berry and Wiseman have proposed an optimal
phase-squeezed state $|\Psi\rangle_{\text{BW}}=\sum_{n}c_{n}|s,
n\rangle$ with the probability amplitude~\cite{Berry}
\begin{equation}
c_{n}=\frac{1}{\sqrt{s+1}}\cos\left(\frac{n\pi}{2s+2}\right).
\label{opPSS}
\end{equation}
Obviously, the BW state has a symmetric probability distribution
about $n=0$ because of $c_{-n}=c_n$, which in turn leads to the
$x$-polarized mean spin $\langle
\vec{S}\rangle=(\langle\hat{S}_{x}\rangle, 0 ,0)$. As shown in
Fig.~\ref{fig1}, one can also find that spin squeezing of the BW
state is along $y$-axis and the anti-squeezing along $z$-axis, i.e.,
\begin{equation}
(\Delta\hat{S}_{y})^{2}<s/2\hskip 5pt\text{ and }(\Delta
\hat{S}_{z})^{2}>s/2. \label{PSCondi}
\end{equation}%
Here, the value $s/2$ ($=N/4$) denotes the standard quantum limit
(SQL).

The phase-squeezed state is a useful resource to implement
Heisenberg-limit metrology~\cite{Berry}. However, it should be
pointed out that the PSS is not an input of linear interferometer,
but a state after the first beam splitter. In other words,
Eq.~(\ref{MZI}) should be rewritten as
$|\Psi\rangle_{\text{out}}=e^{-i\frac{\pi}{2}\hat{S}_{x}}e^{i\Phi
\hat{S}_{z}}|\Psi\rangle_{\text{pss}}$. Inserting it into
Eq.~(\ref{sensitivity}), we obtain the best sensitivity $(\Delta
\Phi)_{0}=(\Delta\hat{S}_{y})_{\text{pss}}/|\langle\hat{S}_{x}\rangle_{\text{pss}}|$
and hence the squeezing parameter
$\zeta_{\text{S}}=\sqrt{N}(\Delta\Phi)_{0}$, where
$\langle\hat{S}_{x}\rangle_{\text{pss}}$ and
$(\Delta\hat{S}_{y})_{\text{pss}}$ denote the expectation value and
the mean-square root of the variance with respect to $|\Psi
\rangle_{\text{pss}}$. As an optimal PSS, the BW state gives rise to
the Heinsenberg-limited sensitivity $(\Delta\Phi)_{0}\sim
2\sqrt{2}/N$ and thus $\zeta_{\text{S}}\sim 2\sqrt{2}N^{-1/2}$ for
large enough particle number ($N>10^2$)~\cite{scalingBW}.

In order to prepare the PSS, we now consider atomic Bose-Einstein
condensate, which is tightly confined in a double
well~\cite{Shin,Jo,Albiez,BVHall,Folling}. By introducing two
bosonic operators $\hat{a}_{\text{L(R)}}$ for left (right)
mode-function $\varphi_{\text{L(R)}}$ and the Schwinger's angular
momentum operators as Eq.~(\ref{MZI}), quantum dynamics of the BEC
system can be described by an effective spin Hamiltonian (with
$\hbar=1$) \cite{Milburn,Smerzi,Cirac,Steel,Javanaine,Vardi01}:
\begin{equation}
\hat{H}_{\gamma}=\delta\hat{S}_{z}+\Omega\hat{S}_{\gamma}+\chi
\hat{S}_{z}^{2},\label{H}
\end{equation}%
where $\delta$ denotes the potential bias, $\Omega$ the amplitude of
Josephson coupling, and $\chi$ the s-wave interaction strength. In
the second term of the Hamiltonian~(\ref{H}), an equatorial spin
operator
$\hat{S}_{\gamma}=\hat{S}_{x}\cos\gamma+\hat{S}_{y}\sin\gamma$ is
introduced to simulate Josephson tunneling of the atoms with Rabi
frequency $\Omega e^{i\gamma}$, where both the amplitude $\Omega$
and the phase $\gamma$ are tunable~\cite{Gross,Riedel}. The third
term, arising from inter- and intraspecies nonlinear interaction,
gives rise to the OAT-induced spin squeezing and many-partite
entanglement~\cite{Kitagawa,Sorensen,Wang,Jin09,Molmer,Rey08}.

Following original spin-squeezing scheme proposed by Kitagawa and
Ueda~\cite{Kitagawa}, and more recently by Riedel {\it et
al.}~\cite{Riedel}, we assume that a $\pi/2$-pulse is applied to the
BEC to prepare the initial states $|s, \pm
s\rangle_x=e^{-i\frac{\pi}{2}\hat{S}_{y}}|s, \pm s\rangle$, which
are coherent spin state~\cite{CSS}:
\begin{equation}
|\theta,
\phi\rangle=\exp[-i\theta(\hat{S}_{y}\cos\phi-\hat{S}_{x}\sin\phi)]|s,
s\rangle, \label{CSS}
\end{equation}%
with polar angle $\theta=\pi/2$ and azimuth angle $\phi=0$ or $\pi$.
During the pulse, the two-mode functions $\varphi_{\text{L,R}}$
merge with each other such that $\delta\simeq\chi\simeq0$ and hence
the Hamiltonian $\hat{H}_{\pi/2}\simeq\Omega\hat{S}_{y}$, where the
phase and the duration of the coupling are adjusted as
$\gamma=\Omega t_{\pi/2}=\pi/2$ with
$\Omega>\!>N\chi$~\cite{Gross,Riedel}. Next, the system evolves
under the Hamiltonian
$\hat{H}_{0}=\Omega\hat{S}_{x}+\chi\hat{S}_{z}^{2}$ for a duration
$t_{\min}$. Nonzero $\chi$ can be realized by splitting the two-mode
BEC that allows for a suitable tunneling with the amplitude $\Omega$
and the phase $\gamma=0$ (or $\pi$), which is fulfilled by applying
a second pulse~\cite{Riedel}.

The interplay of the linear tunneling and the nonlinear interaction
results in dynamical generation of Gaussian number-squeezed
state~\cite{Law,Jin07,Jin08,Jin10} and Schr\"{o}dinger's cat state
(i.e., the N00N state)~\cite{Savage99,Raghavan,mpe2,Hines,Yi,MZI}.
Here, we show that it is also possible to create phase-squeezed
state in the two-mode BEC. To understand how it works, we consider
the Hamiltonian $\hat{H}_{0}$ with time-independent $\chi$ and
$\Omega$. For a fixed particle number $N$ ($=2s$), the system's
energy is conserved, i.e., $d\langle\hat{H}_{0}\rangle/dt=0$, which
yields
$\langle\hat{S}_{z}^{2}(t)\rangle=C-\Omega\langle\hat{S}_{x}(t)\rangle/\chi$
with a constant of integral $C=s/2\pm\Omega s/\chi$ for the initial
CSS $|s, \pm s\rangle_x$, that is
\begin{equation}
\langle\hat{S}_{z}^{2}(t)\rangle=\frac{s}{2}\left[1\pm \frac{2\Omega
}{\chi}\left(1\mp\frac{\langle\hat{S}_{x}(t)\rangle}{s}\right)\right],
\label{exact}
\end{equation}
where the upper sign corresponds to $|s, +s\rangle_{x}$ and the
lower sign for $|s, -s\rangle_{x}$. It is interesting to note that
the variance
$(\Delta\hat{S}_{z})^{2}=\langle\hat{S}_{z}^{2}\rangle>s/2$ only if
the generated state is symmetric and evolved from $|s,
+s\rangle_{x}$ (or $|s, -s\rangle_{x}$) under $\hat{H}_{0}$ with
$\Omega\chi>0$ ($<0$). Except the so-called
`over-squeezing'~\cite{Shalm}, for which both $(\Delta\hat{S}_z)^2$
and $(\Delta\hat{S}_y)^2$ are larger than the SQL, the enhanced spin
fluctuation of relative atom number implies the appearance of a PSS
with $(\Delta\hat{S}_y)^2<s/2$. For brevity, we will assume that
both $\Omega$ and $\chi$ are real and positive.

\subsection{Direct method to prepare the PSS}

\begin{figure}[tbph]
\begin{center}
\includegraphics[width=3in]{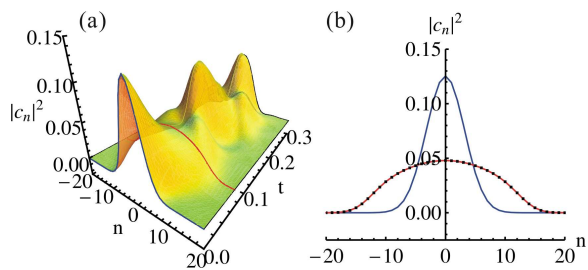}
\caption{(Color online) (a) Probability distribution $|c_{n}|^{2}$
of the spin state evolved from $|s, +s\rangle_x$ as a function of
time $t$ (in units of $\chi^{-1}$); (b) Snapshot of $|c_{n}|^{2}$
for the initial state (blue line) and the PSS at time
$t_{\min}=6.95\times 10^{-2}\chi^{-1}$ (red line with points). Other
parameters: $\Omega/\chi=1.134N$ and $N=2s=40$.} \label{fig2}
\end{center}
\end{figure}

Firstly, let us consider dynamical evolution of the initial CSS $|s,
+s\rangle_{x}$ under the system Hamiltonian $\hat{H}_{0}$. To
distinguish with previous scheme~\cite{mpe2}, we adopt relatively
larger coupling ratio $\Omega/\chi\sim 1.1N$ (see blow
Fig.~\ref{fig4}), which leads to a completely different dynamics due
to the fact that the initial state, i.e., Eq.~(\ref{CSS}) with
$\theta=\pi/2$ and $\phi=0$, now corresponds to a stable fixed point
in phase space. To confirm it, we plot time evolution of probability
distribution $|c_n(t)|^2$. As shown in Fig.~\ref{fig2}(a), the
generated state $|\Psi(t)\rangle$ shows symmetric probability
distribution at any time $t$. For large $N$, the initial CSS $|s,
+s\rangle_x$ is a Gaussian~\cite{Imamoglu}, namely
\begin{equation}
|c_n(0)|^2=\frac{1}{2^{2s}}{\binom{2s}{s+n}}\simeq\frac{1}{\sqrt{\pi
s}}e^{-n^2/s}, \label{iniS}
\end{equation}
with its width determined by the variance
$\Delta\hat{S}_z=\sqrt{s/2}$ (see blue line of Fig.~\ref{fig2}(b),
also Ref.~\cite{Jin10}). After a certain duration
$t_{\min}\simeq\log_2(N)/(2N\chi)$, the initial state will evolve
into a phase-squeezed state, which can be inferred from an increased
width of probability distribution $\Delta\hat{S}_z>\sqrt{s/2}$ (red
line of Fig.~\ref{fig2}). Such a result can be explained from
Eq.~(\ref{exact}). Indeed, we can prove that the mean spin of
$|\Psi(t)\rangle$ aligns with the $x$-axis and its length
$\langle\hat{S}_x(t)\rangle<s$. Therefore, from Eq.~(\ref{exact}) we
have $(\Delta\hat{S}_z)^2>s/2$ and hence $(\Delta\hat{S}_y)^2<s/2$,
except for the over-squeezing.

\begin{figure}[tbph]
\begin{center}
\includegraphics[width=3in]{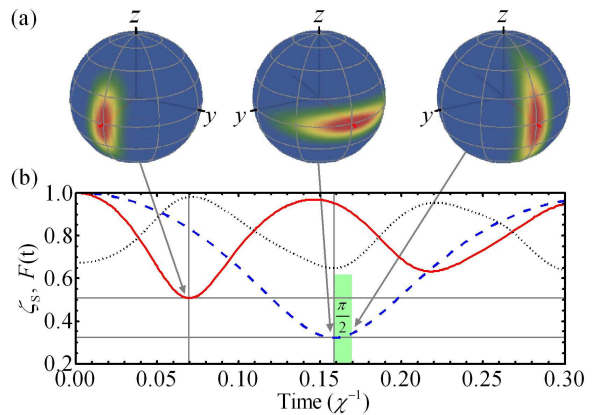}
\caption{(Color online) (a) Quasi-probability distribution
$Q(\theta, \phi)$ of the dynamical generated PSS (left), the NSS
(middle), and the PSS: $e^{i(\pi/2)\hat{S}_{x}}|\Psi
\rangle_{\text{nss}}$ (right); (b) The squeezing parameter
$\zeta_{\text{S}}$ (solid red) and the Fidelity $F(t)$ (dotted) as a
function of time $t$ (in units of $\chi^{-1}$) for the initial
states $|s, +s\rangle_x$ with $s=N/2=20$ and $\Omega/\chi=1.134N$.
The dashed blue line corresponds to evolution of $\zeta_{\text{S}}$
from the initial CSS $|s, -s\rangle_x$, with $s=20$ and
$\Omega/\chi=2.664$.} \label{fig3}
\end{center}
\end{figure}

The appearance of the PSS can be illuminated by calculating
quasi-probability distribution $Q(\theta, \phi)=|\langle\theta,
\phi|\Psi(t)\rangle|^2$ on the Bloch sphere~\cite{Kitagawa}, where
the CSS $|\theta, \phi\rangle$ is defined by Eq.~(\ref{CSS}), with
the polar angle $0\leq\theta\leq\pi$ and the azimuth angle
$0\leq\phi\leq2\pi$. As depicted in the left plot of
Fig.~\ref{fig3}(a), an elliptic shape of the $Q$ function is found
for the spin state $|\Psi(t)\rangle$ at $t_{\min}$, which shows the
squeezing along $\hat{S}_{y}$ and the anti-squeezing along
$\hat{S}_{z}$, similar with that of the BW state. In
Fig.~\ref{fig3}(b), we plot time evolution of the squeezing
parameter $\zeta_{\text{S}}$ (solid red line) and the fidelity
between the prepared state with the BW state,
$F(t)=|_{\text{BW}}\!\langle\Psi|\Psi(t)\rangle|^2$ (dotted line).
For instance at $t=0$, the fidelity is given by
\begin{equation}
F(0)\simeq\frac{2(\pi
s)^{1/2}}{s+1}\exp\left[-\frac{s}{4}\left(\frac{\pi}{1+s}\right)^2\right],
\label{fidelity}
\end{equation}
where we used Eq.~(\ref{opPSS}) and the approximate result of
Eq.~(\ref{iniS}). For the case $N=2s=40$ and $\Omega/\chi=1.134N$,
our analytical result predicts $F(0)\sim0.675$, coincident with
exact numerical simulation $0.672$. From Fig.~\ref{fig3}(b), we find
that the generated PSS at time $t_{\min}=6.95\times
10^{-2}\chi^{-1}$ shows a high fidelity $F=0.982$ and a local
minimum of the squeezing parameter $\zeta_{\text{S}}=0.507$
($-2.949$ dB), which is the optimal squeezing that this scheme can
reach for $N=40$.

It is necessary to investigate the optimal coupling ratio
$\Omega/\chi$ and the minimal value of $\zeta_{\text{S}}$ (i.e., the
maximal squeezing) for any finite $N$. As an example, we consider
the exactly solvable case with $N=2$. Analytical results can be
obtained as the following: $\langle\hat{S}_{x}\rangle
=1-\frac{\chi^{2}}{2\mathcal{E}^{2}}\sin^{2}(\mathcal{E}t)$ and
$(\Delta\hat{S}_{y})^{2}=\frac{1}{2}[1-\frac{\Omega\chi
}{\mathcal{E}^{2}}\sin^{2}(\mathcal{E}t)]$, where
$\mathcal{E}=\sqrt{\Omega^{2}+(\chi/2)^{2}}$, denoting half of
level-spacing between the ground state and the second-excited state.
Therefore, we get the maximal squeezing
$\zeta_{\text{S}}^{2}=(4x^{2}+1)/(2x+1)^{2}$ with $x=\Omega/\chi$,
which occurs at times $t_{k}=(k+\frac{1}{2})\pi/\mathcal{E}$ for an
integer $k=0$, $1$, $\cdots$. Minimizing $\zeta_{\text{S}}^{2}$ with
respect to $x$, we further obtain the optimal coupling ratio
$\Omega/\chi=1/2$ and thus the best squeezing
$\zeta_{\text{S}}=1/\sqrt{2}$. For three-particle case,
$\langle\hat{S}_{x}\rangle
=\frac{3}{2}[1-\frac{\chi^{2}}{\mathcal{E}^{2}}\sin^{2}(\mathcal{E}t)]$
and $(\Delta\hat{S}_{y})^{2}=\frac{3}{4}[1-\frac{2\chi(\Omega-\chi)
}{\mathcal{E}^{2}}\sin^{2}(\mathcal{E}t)]$, where
$\mathcal{E}=\sqrt{\Omega^{2}-\Omega\chi+\chi^{2}}$. The maximal
squeezing also occurs at times $t_{k}$, with
\begin{eqnarray}
\zeta_{\text{S}}^{2}=\frac{(x^{2}-x+1)(x^{2}-3x+3)}{x^{2}(x-1)^{2}},
\end{eqnarray}
from which we find that the optimal squeezing
$\zeta_{\text{S}}=0.762$ is attainable for the coupling ratio
$x=\Omega/\chi=2.794$ ($\sim 0.931N$ with $N=3$).
Counter-intuitively, a slightly weaker spin squeezing is obtained in
a comparison with previous $N=2$ case. This is because the generated
PSS in the two-particle system is also maximally-entangled N00N
state, which results in the Heisenberg-limited sensitivity
$(\Delta\Phi)_{0}=1/N$ and the optimal squeezing
$\zeta_{\text{S}}=1/\sqrt{N}$ (with $N=2$).

The optimal squeezing for $N>3$ has to be determined numerically.
Due to computation limit, only finite particle number $2\leq
N\leq200$ are shown in Fig.~\ref{fig4}. Our numerical simulations
show that the optimal squeezing with $\zeta_{\text{S}}\sim
N^{-0.21}$ can be obtained for the coupling ratio $\Omega/\chi\sim
1.1N$ (see balls of Fig.~\ref{fig4}). Although the achievable
squeezing is worse than that of the BW state (empty circles)
\cite{scalingBW}, it is still useful to achieve the sub-shot noise
in the phase estimation due to $\zeta_{\text{S}}<1$.

\subsection{Indirect method to prepare the PSS via a rotation of the NSS}

Next, we consider dynamical evolution of another initial CSS $|s,
-s\rangle_x$ under governed by the Hamiltonian $\hat{H}_0$.
According to Law et al.~\cite{Law}, it is possible to generate
dynamically a Gaussian number-squeezed state, which shows the
squeezing along $\hat{S}_z$ and the anti-squeezing along $\hat{S}_y$
(see middle plot of Fig.~\ref{fig3}(a)). During time evolution, the
generated state $|\Psi(t)\rangle$ is always symmetric around $n=0$.
More specially, the probability amplitudes obey $c_{-n}(t)=c_{n}(t)$
for even $N$ case, or $c_{-n}(t)=-c_{n}(t)$ for odd $N$ case,
leading to the $x$-polarized mean spin~\cite{Jin07}. Previously, the
optimal coupling ratio $\Omega/\chi$ as a function of $N$ has been
obtained by minimizing the reduced variance over the SQL:
$\xi_{\text{N}}^{2}=2(\Delta\hat{S}_{z})^{2}/s$, which was referred
as the number-squeezing parameter~\cite{Jo,Esteve,Raghavan} and is
measurable by a detection of the phase coherence
$g^{(1)}=|\langle\hat{S}_{x}\rangle|/s$. For the symmetric state,
Eq.~(\ref{exact}) gives us an exact relationship between
$\xi_{\text{N}}^{2}$ and $g^{(1)}$~\cite{Jin08,Jin10}.

\begin{figure}[phtb]
\begin{center}
\includegraphics[width=3in]{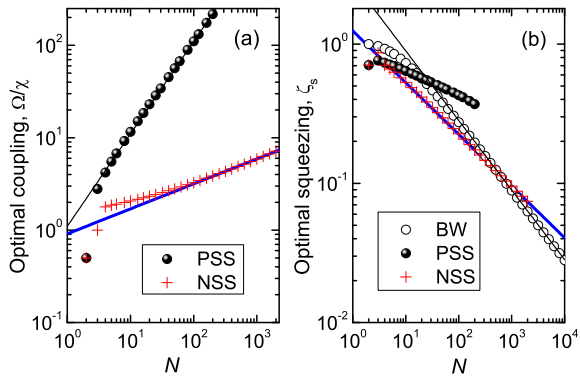}
\caption{(Color online) (a) The optimal coupling ratio $\Omega/\chi$
as a function of $N$ for the PSS (balls) and the NSS (red crosses),
which fit with $1.1N$ (thin solid) and $0.9114N^{0.2704}$ (thick
blue line), respectively; (b) Power rule of the optimal squeezing
$\zeta_{\text{S}}$ for the BW state (empty circles), the PSS
(balls), and the NSS (red crosses). For large $N$ ($>10^2$),
$\zeta_{\text{S}}$ of the BW state and the NSS fit well with
$2\sqrt{2}N^{-1/2}$ (thin solid line) and $1.2432 N^{-0.3715}$
(thick blue line), respectively.} \label{fig4}
\end{center}
\end{figure}

To confirm whether the prepared NSS is useful for metrology or not,
we have to investigate the squeezing parameter
$\zeta_{\text{S}}=\xi_{\text{N}}/g^{(1)}$~\cite{Esteve}. As shown by
dashed blue line of Fig.~\ref{fig3}(b), one can find that the
prepared state $|\Psi(t)\rangle$ at time $t_{\min}=0.159\chi^{-1}$
shows the optimal squeezing $\zeta_{\text{S}}=0.322$ (about $-5$dB)
for the case $N=40$ and $\Omega/\chi=2.664$. In Fig.~\ref{fig4}, we
investigate power rules of the optimal coupling ratio $\Omega/\chi$
and the best squeezing $\zeta_{\text{S}}$. Our numerical results
(the red crosses) show that $\Omega/\chi\sim 0.91N^{0.27}$ and
$\zeta_{\text{S}}\sim1.24N^{-0.37}$. Such a scaling is worse than
that of the BW state $\zeta_{\text{S}}\sim N^{-0.5}$. However, as an
input state of linear interferometer, it is sufficient to achieve
the sub-shot-noise limited metrology.

Before closing, it should be mentioned that the Gaussian
number-squeezed state has been prepared in several
experiments~\cite{Orzel,Strabley,Esteve,Chuu,Jo,Vuletic}, which
exhibits sub-Poissonian atom number distribution~\cite{Orzel,Chuu}
and relatively longer coherence time compared with that of the
product CSS~\cite{Jo}. These features manifest it as a promising
candidate for quantum metrology~\cite{Uys,Huang}. Recently, Uys and
Meystre have found that the NSS gives rise to the Heisenberg-limited
phase sensitivity~\cite{Uys} (see also Ref.~\cite{Huang}). If a
Gaussian NSS is fed into the interferometer as Eq.~(\ref{MZI}), the
action of the input beam splitter $e^{i(\pi/2)\hat{S}_{x}}$,
equivalent with a $\pi/2$-pulse, rotates the prepared NSS about
negative $x$-axis by an angle $\pi/2$, which results in a PSS:
$|\Psi\rangle_{\text{pss}}=e^{i(\pi
/2)\hat{S}_{x}}|\Psi\rangle_{\text{nss}}$, as shown by right plot of
Fig.~\ref{fig3}(a). This is another method to generate
phase-squeezed state~\cite{Smerzi06,Grond}.

\section{Conclusion}

In summary, we have demonstrated that it is possible to generate
phase-squeezed state in a two-mode Bose-Einstein condensate.
Considering a symmetric input state of linear interferometer, we
show that it is easy to derive the best sensitivity and its relation
with the squeezing parameter $\zeta_{\text{S}}$. As an example, we
have discussed general features of the Berry and Wiseman's state:
the $x$-polarized mean spin and the Heisenberg-limited phase
sensitivity. Two experimentally realizable schemes are proposed to
generate the phase-squeezed states. We find that it can be prepared
from an initial coherent spin state $|s, +s\rangle_x$ without any
rotation. Alternatively, starting with $|s, -s\rangle_x$, a
number-squeezed state is prepared and then transformed into
phase-squeezed state via a $\pi/2$-pulse. The optimal coupling and
the maximal squeezing are determined based upon a wide range of
numerical simulations. Our results show that the generated
phase-squeezed states can achieve sub-shot-noise-limited quantum
metrology.

\section*{Acknowledgments}

We thank Prof. L. You and Dr. Y. C. Liu for helpful discussion. This
work is supported by the NSFC (Contract No.~10804007). One of the
authors (Y. An) is partially supported by National Innovation
Experiment Program for University Students (BJTU No.~1070006).



\begin{thebibliography}{99}
\bibitem{scalingBW} Analytically, the authors of Ref.~\cite{Berry} have
derived the best sensitivity $(\Delta \Phi)_{0}=\tan[\pi/(N+2)]\sim
\pi/N$ for the optimal PSS, Eq.~(\ref{opPSS}). Our numerical
simulations show $(\Delta \Phi)_{0}\sim 2\sqrt{2}/N$ and hence
$\zeta_{\text{S}}\sim 2\sqrt{2}/N^{1/2}$ [see the empty circles of
Fig.~4(b)].
\end{thebibliography}
\end{document}